\documentclass[twocolumn,reprint,superscriptaddress,amsmath,amssymb,aps,nofootinbib]{revtex4-1}

\usepackage{graphicx}
\usepackage{dcolumn}
\usepackage{bm}

\usepackage{hyperref}
\usepackage[utf8]{inputenc}
\usepackage[dvipsnames]{xcolor}
\usepackage[normalem]{ulem}
\usepackage{xspace}
\usepackage{fontawesome} 
\usepackage{multirow}

\usepackage[dvipsnames]{xcolor}
\usepackage[utf8]{inputenc}
\hypersetup{
    colorlinks=true,
    citecolor = blue,
    linkcolor=red,
    filecolor=magenta,      
    urlcolor=magenta,
}

\newcommand{\nucleus}{\mathcal{N}}

\newcommand{\UniNA}{%
Universit\`a degli Studi di Napoli ``Federico II'', Complesso Univ. Monte S. Angelo, I-80126 Napoli, Italy, and INFN - Sezione di Napoli,
Complesso Univ. Monte S. Angelo, I-80126 Napoli, Italy}
\newcommand{\SSM}{%
Scuola Superiore Meridionale, Universit\`a di Napoli Federico II, Largo San Marcellino 10, 80138 Napoli, Italy}
\newcommand{\UniBa}{%
Dipartimento Interateneo di Fisica ``Michelangelo Merlin'', Universit\`a degli Studi di Bari, Via Amendola 173, 70126 Bari, Italy}
\newcommand{\INFNBa}{%
Istituto Nazionale di Fisica Nucleare, Sezione di Bari, Via Orabona 4, 70126 Bari, Italy}

\begin{document}

\title{Primordial Black Hole Dark Matter evaporating on the Neutrino Floor}%

\author{Roberta Calabrese}
\affiliation{\UniNA}%
\author{Damiano F.G. Fiorillo}%
\affiliation{\UniNA}%
\author{Gennaro Miele}
\affiliation{\UniNA}
\affiliation{\SSM}
\author{Stefano Morisi}%
\affiliation{\UniNA}
\author{Antonio Palazzo}
\thanks{Corresponding author}
\email{email: antonio.palazzo@ba.infn.it}
\affiliation{\UniBa}
\affiliation{\INFNBa}


\begin{abstract}

Primordial black holes (PBHs) hypothetically generated in the first instants of life of the Universe are potential
dark matter (DM) candidates. Focusing on PBHs masses  in the range $[5 \times10^{14} - 5 \times 10^{15}]$g, we point out that the neutrinos emitted by PBHs evaporation can interact through the coherent elastic neutrino nucleus scattering (CE$\nu$NS) producing an observable signal in multi-ton DM direct detection experiments. We show that with the high exposures envisaged for the next-generation facilities, it will be possible to set bounds on the fraction of DM composed by PBHs improving the existing neutrino limits obtained with Super-Kamiokande. We also quantify to what extent a signal originating from a small fraction of DM in the form of PBHs would modify the so-called ``neutrino floor'', the well-known barrier towards  detection of weakly interacting massive particles (WIMPs) as the dominant DM component. 

\end{abstract}

\maketitle

{\bf \em Introduction.} The identity of dark matter (DM) is one of the most puzzling mysteries in 
contemporary astroparticle physics and cosmology. In spite of enormous efforts, no uncontroversial
non-gravitational signal of DM has emerged so far. In this context, 
the recent detection of gravitational waves from binary black hole mergers by 
LIGO/Virgo~\cite{Abbott:2020tfl,LIGOScientific:2018jsj}  has strongly revamped  the attention~\cite{Bird:2016dcv,Clesse:2016vqa,Sasaki:2016jop,Blinnikov:2016bxu,Nakama:2016gzw} 
towards the hypothesis that DM may be composed of primordial black holes (PBHs). 
As first recognized in the seventies, these objects can be generated in the early
Universe from the collapse of large overdensities~\cite{Zeldovich:1967lct,Hawking:1971ei,Carr:1974nx,Hawking:1974rv,Carr:1975qj,Chapline:1975ojl,Khlopov:1985jw},
 and may constitute a fraction of the observed amount of
DM~\cite{Carr:1974nx,Chapline:1975ojl} (see~\cite{Sasaki:2018dmp,Carr:2020gox,Green:2020jor,Carr:2020xqk,Villanueva-Domingo:2021spv} for recent reviews).
PBHs emit Hawking radiation~\cite{Hawking:1974sw}, and for
large enough masses ($M_{\rm PBH}\gtrsim 5 \times 10^{14}\rm g$), 
have a lifetime longer than the age of the Universe.
The evaporation process can give
rise to observable signals. In fact, bounds (present or prospective) on PBHs have been
obtained from X-rays~\cite{Ballesteros:2019exr,Iguaz:2021irx}, $\gamma$-rays~\cite{Carr:2016hva,Arbey:2019vqx,Laha:2020ivk,Ballesteros:2019exr,Coogan:2020tuf,Ray:2021mxu},
511 keV $\gamma$-ray line from galactic center~\cite{Laha:2019ssq,Dasgupta:2019cae,Keith:2021guq}, 
galactic $e^\pm$~\cite{Boudaud:2018hqb}, 
cosmic microwave background (CMB)~\cite{Poulin:2016anj,Clark:2016nst,Poulter:2019ooo}, 
radio signals from inverse Compton scattering on CMB photons~\cite{Dutta:2020lqc} and
synchrotron radiation~\cite{Chan:2020zry},
and heating of  the interstellar medium~\cite{Laha:2020vhg,Kim:2020ngi}.
The possibility to constrain PBHs using the emitted neutrinos was discussed
 long ago in~\cite{Halzen:1995hu,Bugaev:2000bz,Bugaev:2002yt}. More recently, limits on
PBHs have been obtained in~\cite{Dasgupta:2019cae}
exploiting the null searches of the diffuse supernova neutrino background (DSNB) performed by Super-Kamiokande~\cite{Bays:2011si}.
Also, prospective bounds from the experiment JUNO have been discussed in~\cite{Wang:2020uvi}.

In this {\em Letter}, we entertain a novel possibility, never addressed before in the literature, proposing 
to detect the emitted neutrinos from PBHs by coherent elastic neutrino nucleus scattering (CE$\nu$NS).
It is only recently that the CE$\nu$NS process, predicted long time ago~\cite{Freedman:1973yd}, has been
successfully observed by COHERENT~\cite{Akimov:2017ade}, where a few kilograms detector
was exposed to an intense neutrino flux of artificial origin.
The very same process involving neutrinos of natural origin, such as the solar, 
DSNB and atmospheric ones, constitutes an irreducible background~\cite{Cabrera:1984rr,Drukier:1986tm,Monroe:2007xp,Vergados:2008jp,Strigari:2009bq,Gutlein:2010tq,Billard:2013qya}
(forming the so-called ``neutrino floor''~\cite{Billard:2013qya}) towards
detection and identification of WIMPs as DM candidates in next-generation direct detection experiments.
Here we show that neutrinos from PBHs with masses in the range $[5 \times10^{14} , 5 \times 10^{15}]$g,
which emit neutrinos with peak energy between 10 MeV and 100 MeV,
may emerge as a signal on top of such a familiar background.
As a consequence, it is possible to set prospective bounds on the PBHs fraction $f_{\rm PBH}$ of DM 
in this mass range. As an interesting byproduct of our study, 
we show how the neutrino floor gets modified by the presence of a hypothetical signal from
PBHs.

\begin{figure}[t!]
\vspace*{-0.7cm}
\hspace*{-0.2cm}
\includegraphics[width=9.5cm]{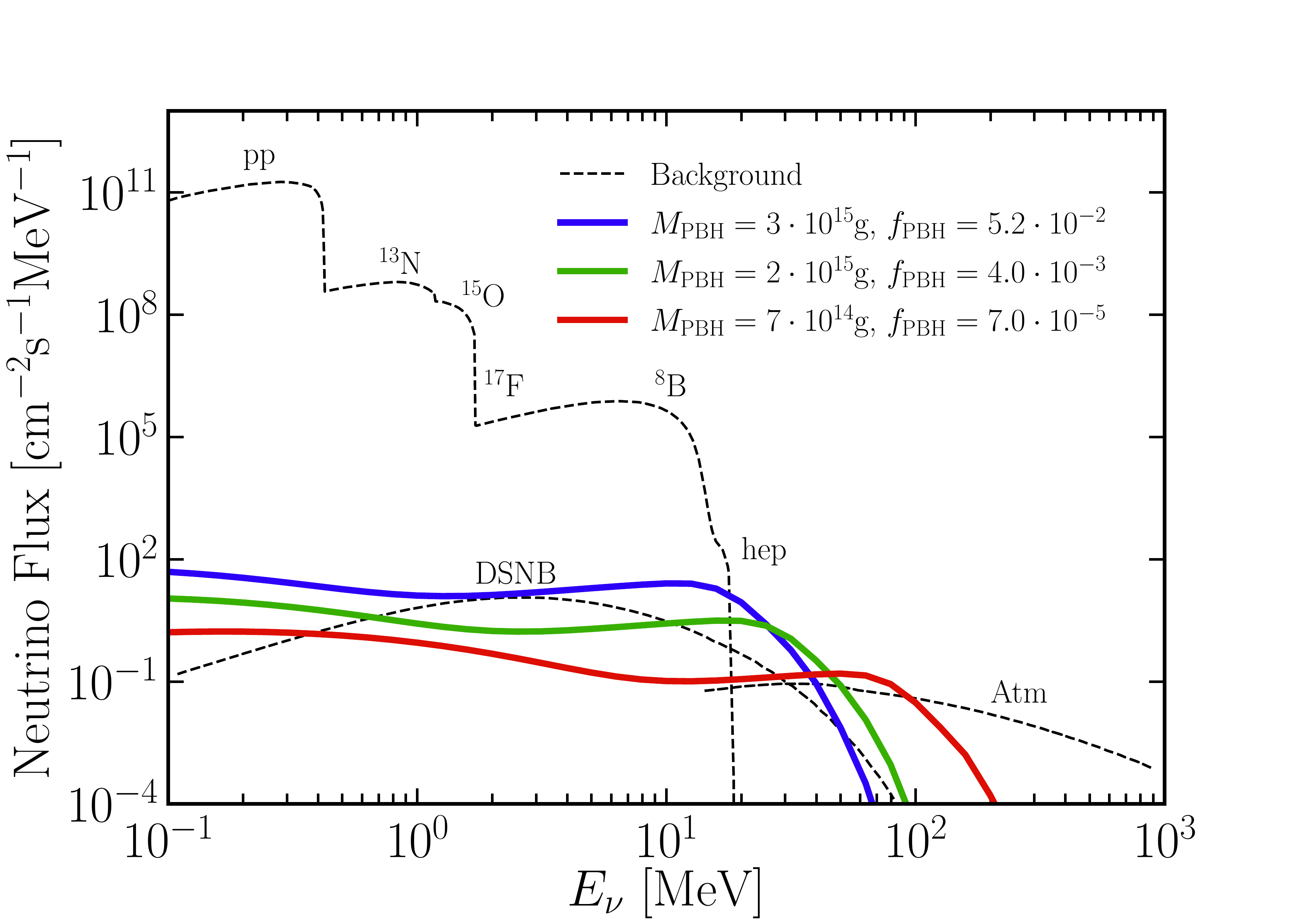}
\vspace*{-0.55cm}
\caption{{\bf Neutrino fluxes from PBHs.}  The black dashed contours represent the
background fluxes originating from solar, DSNB and atmospheric neutrinos. 
The colored solid lines correspond to neutrinos from PBHs evaporation for three representative
values of their mass and fraction of total DM content. These benchmark values lie on the
90\% C.L. exclusion curve obtainable from a liquid xenon experiment with 200 t yr exposure
(corresponding to the black stars in Fig.~\ref{fig:upper_bounds}).
\label{fig:nu_fluxes}}
\end{figure}

{\bf \em Neutrinos emitted by PBHs.} 
A black hole is believed to quantum evaporate~\cite{Hawking:1974sw},
emitting radiation akin to a hot body. For a neutral non-rotating (Schwarzschild) 
black hole, the Hawking temperature is given by~\cite{Hawking:1974sw, Page:1976df,Page:1977um, MacGibbon:1990zk}
\begin{equation}
k_{\rm B} T_{\rm{PBH}} = \frac{\hbar c^3}{8 \pi G_{N} M_{\rm{PBH}}}  \simeq 1.06 \left[\frac{10^{16}\rm{g}}{M_{\rm{PBH}}}\right]\rm{MeV}\,,
\label{eq:PBH temperature}
\end{equation}
where four fundamental constants appear, $k_{\rm B}$ (Boltzmann), $G_N$ (gravitational), $\hbar$
(Planck), and $c$ (speed of light). The differential flux per unit time of emitted particles depends on their spin. For spin 1/2 
particles with mass negligible with respect to $T_{\rm PBH}$ like neutrinos, it is given by
\begin{equation}
\frac{d^2N_\nu}{dE_\nu dt} = \frac{1}{2\pi} \frac{\Gamma_\nu(E_\nu,M_{\rm{PBH}})}{{\exp}\left[{E_\nu}/{k_{\rm B}T_{\rm{PBH}}} \right]+1} \,,
\label{eq:Instanteneus}
\end{equation}
where $E_\nu$ is the neutrino energy and $\Gamma_\nu$ is a graybody 
factor~\cite{Page:1976df,Page:1977um, MacGibbon:1990zk} encoding the 
imprint of the space-time geometry intervening between the event horizon and the asymptotic observer. 
In our analysis, we employ the publicly available {\tt BlackHawk} code~\cite{Arbey:2019mbc} 
for calculating the energy spectra of the emitted neutrinos.
{\tt BlackHawk} provides the primary spectra for all fundamental Standard Model particles
using the Hawking evaporation spectrum in Eq.~(\ref{eq:Instanteneus}).
In addition, the code generates the spectra of secondary neutrinos deriving 
from hadronization of strongly-interacting constituents and decay of unstable particles.
We sum up both kind of spectra in our analysis. In~\cite{Lunardini:2019zob}, it has been shown
that Dirac neutrinos, having twice as many degrees of freedom as the Majorana ones, would
affect the PBHs evaporation making it faster (see also~\cite{Hooper:2019gtx,Masina:2020xhk}).
Although the additional Dirac degrees of freedom are sterile for
the electroweak interactions and not detectable in the CE$\nu$NS process, their existence
can alterate indirectly the emission rate of the active degrees of freedom. For the PBHs masses considered
in our work, this effect can be quantified around $\sim 10\%$~\cite{Lunardini:2019zob}. 
For definiteness, we assume that neutrinos have Majorana nature. We ignore neutrino 
oscillations being irrelevant for the flavor-blind CE$\nu$NS process. We take into account both the contributions coming 
from the galactic and extragalactic PBHs. The galactic differential neutrino flux is given by  
\begin{equation}
\frac{d\phi_\nu^{\rm MW}}{dE_\nu} = \int \frac{d\Omega}{4\pi}   \frac{d^2N}{dE_\nu dt} \int dl\,\frac{f_{\rm PBH}\,\rho_{\rm MW}\left[r(l,\psi)\right]}{M_{\rm{PBH}}}\,,
\label{eq:F_MW}
\end{equation}
where $\rho_{\rm MW}(r)$ is the DM density of the Milky Way (MW), $r$ denotes the galactocentric distance 
\begin{equation}
r(l,\psi) = \sqrt{r_\odot^2 -2 l r_\odot \cos \psi + l^2}\,,
\end{equation}
with $r_\odot$ being the distance of the Sun from the galactic center, $l$ the line-of-sight distance to the PBH,
$\psi$ the angle between these two directions, and $\Omega$ the solid angle under consideration.
For definiteness, we employ a Navarro-Frenk-White (NFW) profile~\cite{Navarro:1996gj}
\begin{align}
  \rho_{\rm MW}(r) = \rho_\odot \left[ \frac{r}{r_\odot}
  \right]^{-1} \left[
  \frac{1+r_\odot/r_s}{1+r/r_s}
  \right]^{2}\,,
  \label{eq:NFW}
\end{align}
where we take $\rho_{\odot}$ = $ 0.4\, {\rm GeV\,cm^{-3}}$ for the DM density in the solar 
neighborhood, $r_{\odot} = 8.5\, {\rm kpc}$, and $r_{s}=20\, {\rm kpc}$
for the scale radius.  We stress that the value $\rho_{\odot}$ = $ 0.4\, {\rm GeV\,cm^{-3}}$ lies 
at the lower end of its allowed range according to the latest estimates (see for example~\cite{Benito:2019ngh}). 
Therefore, the bounds we are going to derive will be conservative in this respect.
For the extragalactic component, the differential
flux integrated over the full sky is~\cite{Carr:2020gox}
\begin{equation}
	\frac{d\phi_\nu^{\rm EG}}{dE_\nu} =  \int_{t_{min}}^{t_{max}} dt\,\,\left[1+z(t)\right] \frac{f_{\rm{PBH}} \rho_{\rm DM}}{M_{\rm{PBH}}} \frac{d^2N_\nu}{d\tilde{E_\nu} dt}\Bigr\rvert_{\tilde{E_\nu} = [1+z(t)] E_\nu}\,,
	\label{eq:extragalactic contribution}
	\end{equation}
$\rho_{\rm DM}$ being the average DM density of the Universe at  the present epoch, as determined 
by Planck~\cite{Aghanim:2018eyx}.  The time integral runs from  $t_{\rm min}$ set to the era of
matter-radiation equality to $t_{\rm max}$, the minimum between the PBH lifetime and the age of the Universe. 
The overall neutrino flux from the sum of galactic and extragalactic contributions
is plotted in Fig.~\ref{fig:nu_fluxes} for three benchmark values of $M_{\rm{PBH}}$ and $f_{\rm{PBH}}$.
As it will be discussed in the next section, these values are excludable at 90\% C.L. from a xenon experiment 
with 200 t yr exposure, assuming a measurement compatible with the ordinary background. In the same plot, 
we represent the background which is formed from solar~\cite{Vinyoles:2016djt}, DSNB~\cite{Beacom:2010kk} 
and low-energy atmospheric neutrinos~\cite{Battistoni:2005pd,Honda:2011nf}. As expected from 
Eq.~(\ref{eq:PBH temperature}), smaller PBHs masses correspond to harder spectra of the emitted neutrinos
with peak located at $\sim 4.2\, T_{\rm{PBH}}$~\cite{Kohri:1999ex}. We see that PBHs neutrinos can be visible
above the abrupt fall-off of the solar $hep$ neutrinos, where the dominant contribution to the background is
provided by DSNB and atmospheric neutrinos.

\begin{figure}[t!]
\vspace*{-0.7cm}
\hspace*{-0.15cm}
\includegraphics[width=9.6cm]{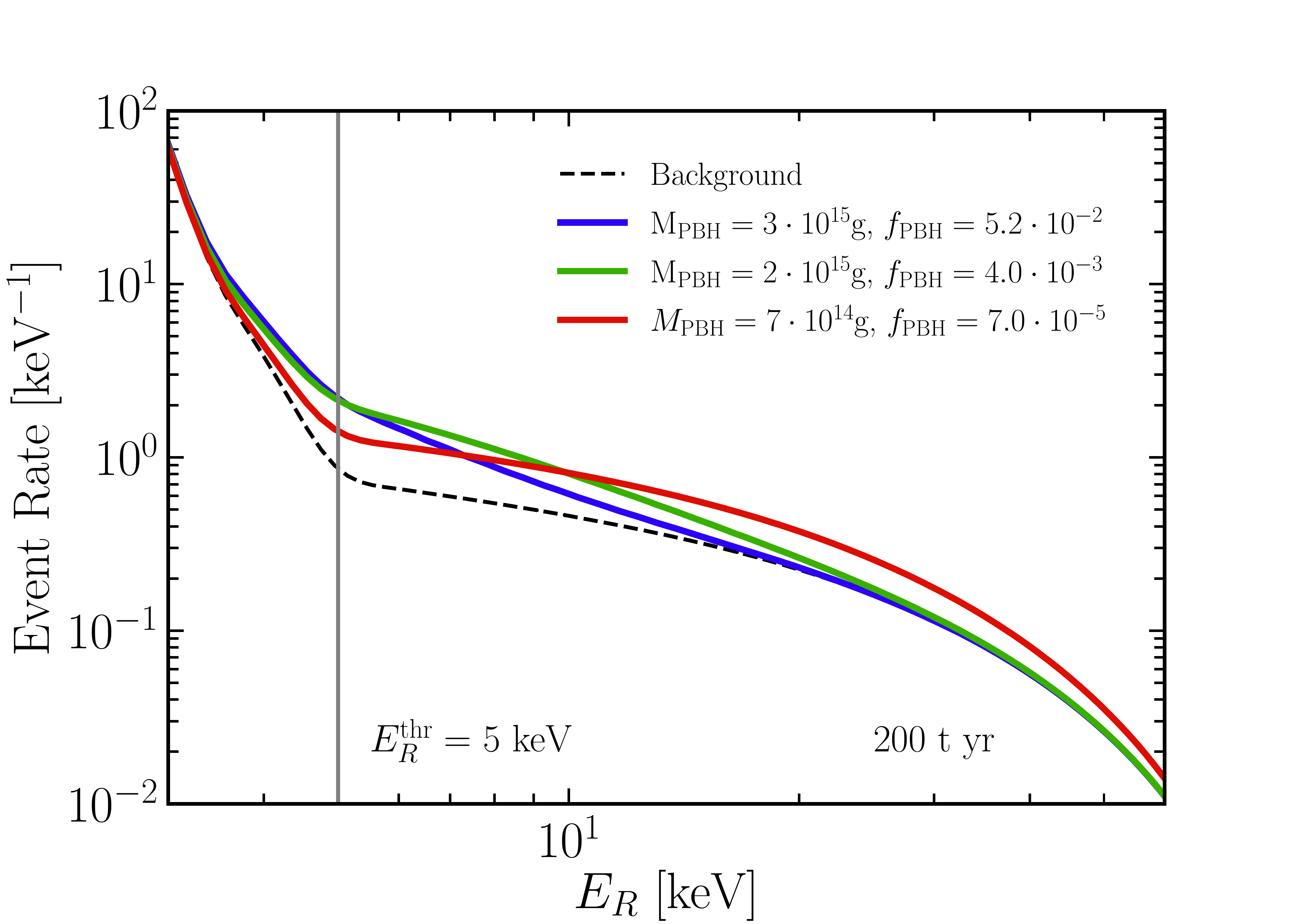}
\vspace*{-0.55cm}
\caption{{\bf Differential neutrino events rate.} The black dashed contour represents
the total background rate (solar + DSNB + atmospheric).
The colored solid lines correspond to neutrinos emitted from PBHs evaporation 
for three representative values of their mass and fraction of total DM content. These
benchmark values  are excludable at the 90\% C.L. by a liquid xenon experiment with 200 t yr exposure
(corresponding to the black stars in Fig.~\ref{fig:upper_bounds}).
The grey vertical line indicates the threshold recoil energy used in the analysis.
\label{fig:event_rate}}
\end{figure}

{\bf \em Coherent scattering of neutrinos.} 
Coherent elastic scattering of a neutrino $\nu$ (or antineutrino $\bar \nu$) on a nucleus $\nucleus$
can occur if $q R \ll 1$, where $q=|\vec{q}|$ is the three-momentum transfer and $R$ is
the nuclear radius~\cite{Freedman:1973yd}.
The differential cross section can be expressed as~\cite{Freedman:1973yd}
\begin{equation}
\label{eq:sigma_nu_N}
   \dfrac{d\sigma_{\nu\text{}\nucleus}}{d E_R}(E_\nu,E_R)= \frac{G_\mathrm F^2 m_\nucleus}{4\pi} Q_w^2 \left(1 - \frac{m_\nucleus E_R}{2 E_\nu^2} \right) F^2(q) \ ,
\end{equation}
where $G_\mathrm F$ is the Fermi constant, $m_\nucleus$ is the nucleus mass, 
$Q_w =\left[N - Z(1 - 4\sin^2\theta_W)\right]$ is the weak vector nuclear charge, 
$Z$ and $N$ are the number of protons and neutrons, $\sin^2\theta_W = 0.231$~\cite{Patrignani:2016xqp} is 
the Weinberg angle, $E_R$ is the nucleus recoil energy, and $E_\nu$ is the neutrino energy.
The recoil energy can assume the maximum value
$E_R^\mathrm{max} = 2E_\nu^2 / (m_\nucleus + 2E_\nu)$.  For the nuclear form factor $F(q)$, encoding
the loss of coherence for $qR > 1$, we employ the Helm parametrization~\cite{Helm:1956zz}
using the recipes provided in~\cite{Lewin:1995rx}. 
The differential rate of events  is given by
\begin{equation}
\label{eq:Event_rate_nu_N}
\begin{aligned}
    \frac{dR_{\nu \nucleus}}{dE_R dt} &= N_T \; \epsilon(E_R) \\
    & \times  \int dE_{\nu} \ \frac{d\sigma_{\nu \nucleus}}{dE_R} \ \frac{d\phi_\nu}{dE_\nu} \ \Theta (E_R^\mathrm{max} - E_R) \ ,
\end{aligned}
\end{equation}
where $N_T$ is the number of target nuclei in the detector, $\epsilon(E_R)$ is the detector efficiency
(assumed to be equal to one), $d\phi_\nu/dE_\nu$ is the differential neutrino flux, and  $\Theta$ is the
Heaviside step function. In Fig.~\ref{fig:event_rate}, we show the differential rate of events
as a function of the recoil energy for the background (dashed line), and for three benchmark values
of the PBHs parameters (the same used in Fig.~\ref{fig:nu_fluxes}). The plots refers to an
exposure of 200 t yr. One can observe that the shape of the spectrum induced by PBHs neutrinos appreciably changes
with the value of the mass of the PBH. In particular, for smaller values of $M_{\rm PBH}$, which
correspond to more energetic neutrino fluxes, the event spectrum is similar to the atmospheric
background. In contrast, for larger values of $M_{\rm PBH}$ with flux peaked at lower energies,
the PBHs event spectrum is quite different with respect to the background. For this reason, in the statistical
treatment presented in the next section we employ a binned likelihood analysis, so as to
exploit the information contained in the spectral shape. 

\begin{figure}[t!]
\vspace*{-0.7cm}
\hspace*{-0.2cm}
\includegraphics[width=9.5cm]{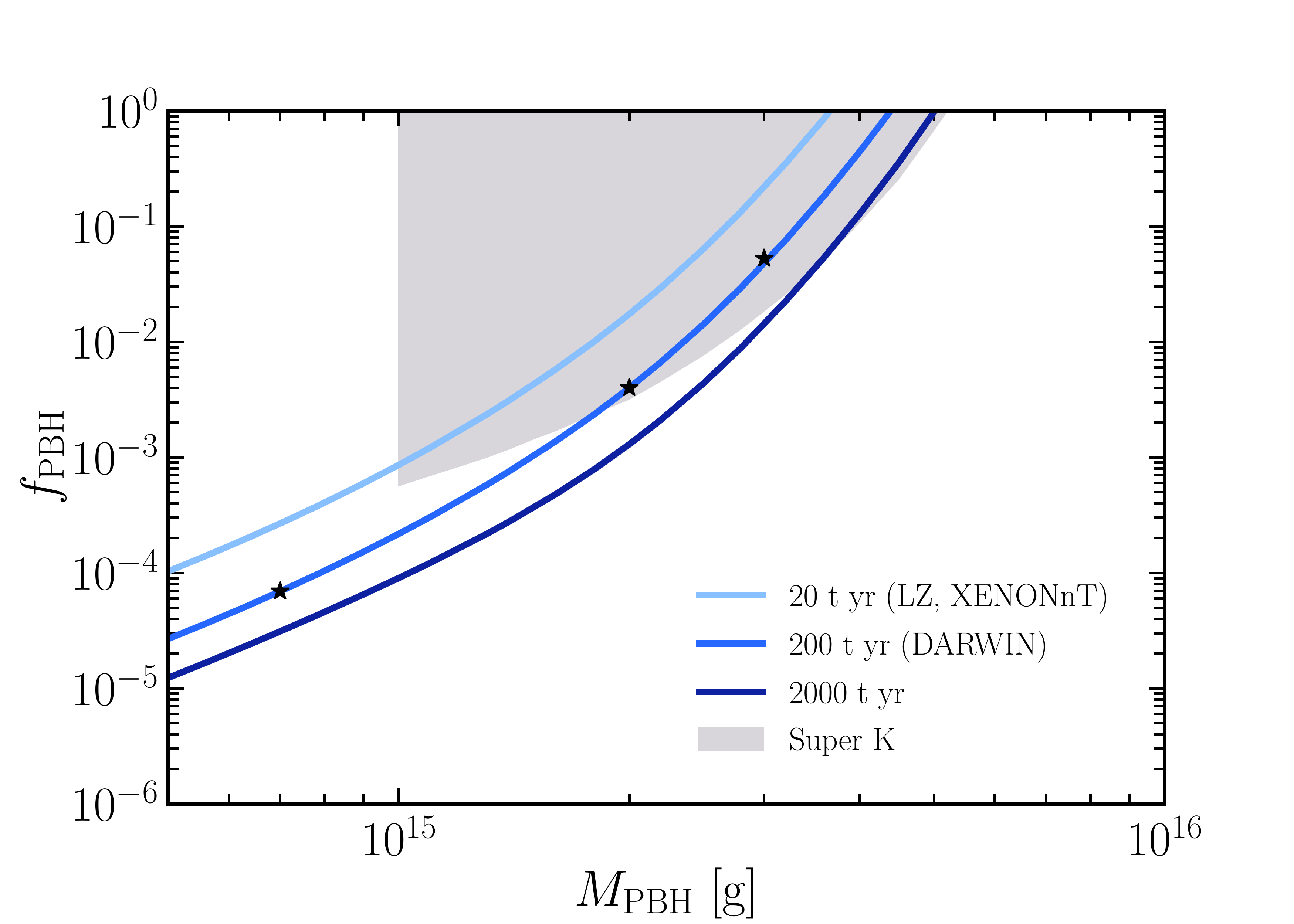}
\vspace*{-0.55cm}
\caption{{\bf Upper bounds on PBHs.}  The exclusion contours, all drawn at the 90\% C.L.
for 1 d.o.f., refer to a xenon experiment with three different exposures.  Two of them correspond 
to the planned experiments (LZ/XENONnT and DARWIN), while the third one refers to an
ideal setup. The black stars indicate the three benchmark points used for Figures \ref{fig:nu_fluxes},
\ref{fig:event_rate} and \ref{fig:nu_floor}.
The shaded grey region represents the upper bounds obtained in~\cite{Dasgupta:2019cae}
from the DSNB searches of Super-Kamiokande~\cite{Bays:2011si}.
\label{fig:upper_bounds}}
\end{figure}

{\bf \em PBHs at next-generation detectors.}  In order to derive prospective upper limits on the
fraction of DM  composed of PBHs, we implement the $\chi^2$ test statistic defined as
\begin{equation}
\label{eq:chi2}
\chi^2 ({\boldsymbol \theta})= \underset{\boldsymbol \alpha}
{\mathrm{min}}{\left[ \chi^2(\boldsymbol \theta,\boldsymbol \alpha)+  (1-\boldsymbol \alpha)^T{\Sigma_{\boldsymbol \alpha}^{-1}}(1-\boldsymbol \alpha)\right]} \ , 
\end{equation}
where ${\boldsymbol \theta}^T=[M_{\rm PBH}, f_{\rm PBH}]$, and 
$\boldsymbol \alpha^T = [\alpha_1, \alpha_1, \alpha_3]$ represent
respectively the vector of the model parameters and that of the nuisance parameters associated to
the normalization of the three backgrounds components (solar $hep$, DSNB, atmospheric) 
with respect to their best theoretical estimates. The uncertainties on the backgrounds are 
encoded in the covariance matrix 
$\Sigma_{\boldsymbol \alpha} =
\rm{diag}(\sigma_{\alpha 1}^2,
\sigma_{\alpha 2}^2,
\sigma_{\alpha 3}^2)$, which
we take diagonal because the three fluxes have completely independent origin.  
For the uncertainties, we have assumed 30\%, 50\% and 20\%, respectively 
for solar $hep$~\cite{Vinyoles:2016djt}, DSNB~\cite{Beacom:2010kk} and atmospheric
neutrinos~\cite{Battistoni:2005pd,Honda:2011nf}. The first term in Eq.~(\ref{eq:chi2}) is defined as 
\begin{equation}
\label{eq:chi2_2}
 \chi^2(\boldsymbol \theta,\boldsymbol \alpha) = -2 \ln{\frac{L_0}{L_1}} \,,
\end{equation}
with likelihoods given by
\begin{equation}
\label{eq:L_0}
L_0 = \prod_i P(x =\overline N^i_{\mathrm{Bck}};\,\lambda = N^i_\mathrm{PBH}(\boldsymbol\theta)+ N^i_\mathrm{Bck}(\boldsymbol\alpha)) \ ,
\end{equation}
and 
\begin{equation}
\label{eq:L_1}
L_1 = \prod_i P(x =  \overline N^i_{\mathrm{Bck}};\, \lambda = \overline N^i_{\mathrm{Bck}}) \ ,
\end{equation}
where $P(x,\lambda)$ is the Poisson distribution, $i$ is the bin index, $\overline N^i_{\mathrm{Bck}}$ is 
the nominal number of background events expected in the $i$-th bin, whereas
$N^i_{\rm{PBH}}$ is the number of PBHs events, and  $N^i_{\rm{Bck}} = \sum_j\alpha_j \overline N^i_{{\rm{Bck}},j}$ 
is the floating background. The second contribution in Eq.~(\ref{eq:chi2}) is a penalty term gauging
the statistical weight of the deviation of the background fluxes from their central values. 
We have neglected other possible backgrounds, in particular those arising from  
electron recoils of solar $pp$ and $^7$Be neutrinos, since as shown in~\cite{Newstead:2020fie}, these
can be effectively suppressed through a statistical discrimination of the photon and ionization signals.
In our statistical analysis, we employ ten bins in the recoil energy window  [$5-50$]~keV.
The choice of the threshold $E_R^{\rm thr} = 5$\,keV is dictated by the position of the sharp cutoff of the $hep$ neutrino
event rate, which in this way has a marginal impact in our results. 
In view of the most recent theoretical  evaluations of the DSNB~\cite{Priya:2017bmm,Horiuchi:2017qja,Moller:2018kpn,Kresse:2020nto}, 
which point towards somewhat larger fluxes of the $\overline \nu_e$ component with respect to previous findings,
as a check, we have increased by a factor of 
two the nominal value~\cite{Beacom:2010kk} of the DSNB flux for both $\nu$ and $\overline\nu$ and for all three flavors.
 We find that the exclusion limits are basically unchanged because in the region of interest the DSNB 
has a subleading role, with the dominant contribution arising from the atmospheric neutrinos.
Figure~\ref{fig:upper_bounds} displays the 90\% C.L. exclusion limits obtainable from a
xenon detector with the three benchmark exposures reported in the legend.
The 20 t yr exposure should be attainable in LZ and XENONnT, while the higher value of 
200 t yr refers to the more ambitious project DARWIN. The third contour corresponds to 2000 t yr, 
which probably may be considered as an ultimate goal for liquid xenon detectors. Such high exposures seem
to be less extreme for Argon based detectors~\cite{Aalseth:2017fik}. In the same plot, for comparison, we report as a
shaded grey region, the upper bounds obtained in~\cite{Dasgupta:2019cae} from the DSNB searches 
of Super-Kamiokande~\cite{Bays:2011si}.%
\footnote{In~\cite{Dasgupta:2019cae}, which considers both spinning and non-spinning PBHs,
the upper limits are reported for masses above $10^{15}$g because rotating black holes evaporate
faster than the non-rotating ones and can contribute to the present DM content for $M_{\rm PBH}\gtrsim 7 \times 10^{15}$g. 
In our work, focused on non-rotating PBHs, we show the limits down to $M_{\rm PBH} = 5 \times 10^{15}$g,
as it is customary in the literature.} 
We see that an improvement of one order of magnitude is attainable with the most high exposures. 
Further amelioration will be possible with a better theoretical knowledge of the atmospheric neutrino fluxes.

{\bf \em Impact of PBHs on the neutrino floor.}  
Neutrinos originating from the Sun, diffuse supernovae and Earth's atmosphere constitute
an irreducible background in DM direct searches~\cite{Cabrera:1984rr,Drukier:1986tm,Monroe:2007xp,Vergados:2008jp,Strigari:2009bq,Gutlein:2010tq}. 
This background gives rise to the so-called  ``neutrino floor'', an ultimate limit to the 
discovery potential in the plane spanned by the WIMP mass $m_\chi$ and the spin-independent WIMP-nucleon cross
section $\sigma_{\chi n}$. Of course, such a limitation holds for any kind of interaction,
and can be generalized to other types of WIMP-nucleus effective field theory operators~\cite{Dent:2016iht,Gelmini:2018ogy},
but is customary to adopt the case of spin-independent interaction as a benchmark. As a matter
of fact, the running experiment  XENON1T~\cite{Aprile:2020thb} is already surfacing the solar $^8$B neutrino background,
which is dominant for WIMP masses around $m_\chi \sim 6$\,GeV. The next-generation facilities
with very high exposures will unavoidably encounter the neutrino floor also for larger WIMP 
masses, where the background from DSNB and atmospheric neutrinos is relevant.
Going below such a floor will require exploiting timing structure of the signal~\cite{Davis:2014ama},
combining different targets~\cite{Ruppin:2014bra}, or using directional information~\cite{Grothaus:2014hja,OHare:2015utx,OHare:2020lva}. Interestingly, the neutrino background can be affected by exotic neutrino
interactions~\cite{Harnik:2012ni,Cerdeno:2016sfi,Papoulias:2018uzy,Suliga:2020jfa}, with
consequent modification of the standard neutrino floor~\cite{Dent:2016wor,Bertuzzo:2017tuf,Gonzalez-Garcia:2018dep,Boehm:2018sux,Chao:2019pyh,Sadhukhan:2020etu}, which can be influenced also by neutrinos originating from decay of massive particles~\cite{Cui:2017ytb,Nikolic:2020fom}.

\begin{figure}[t!]
\vspace*{-0.7cm}
\hspace*{-0.2cm}
\includegraphics[width=9.5cm]{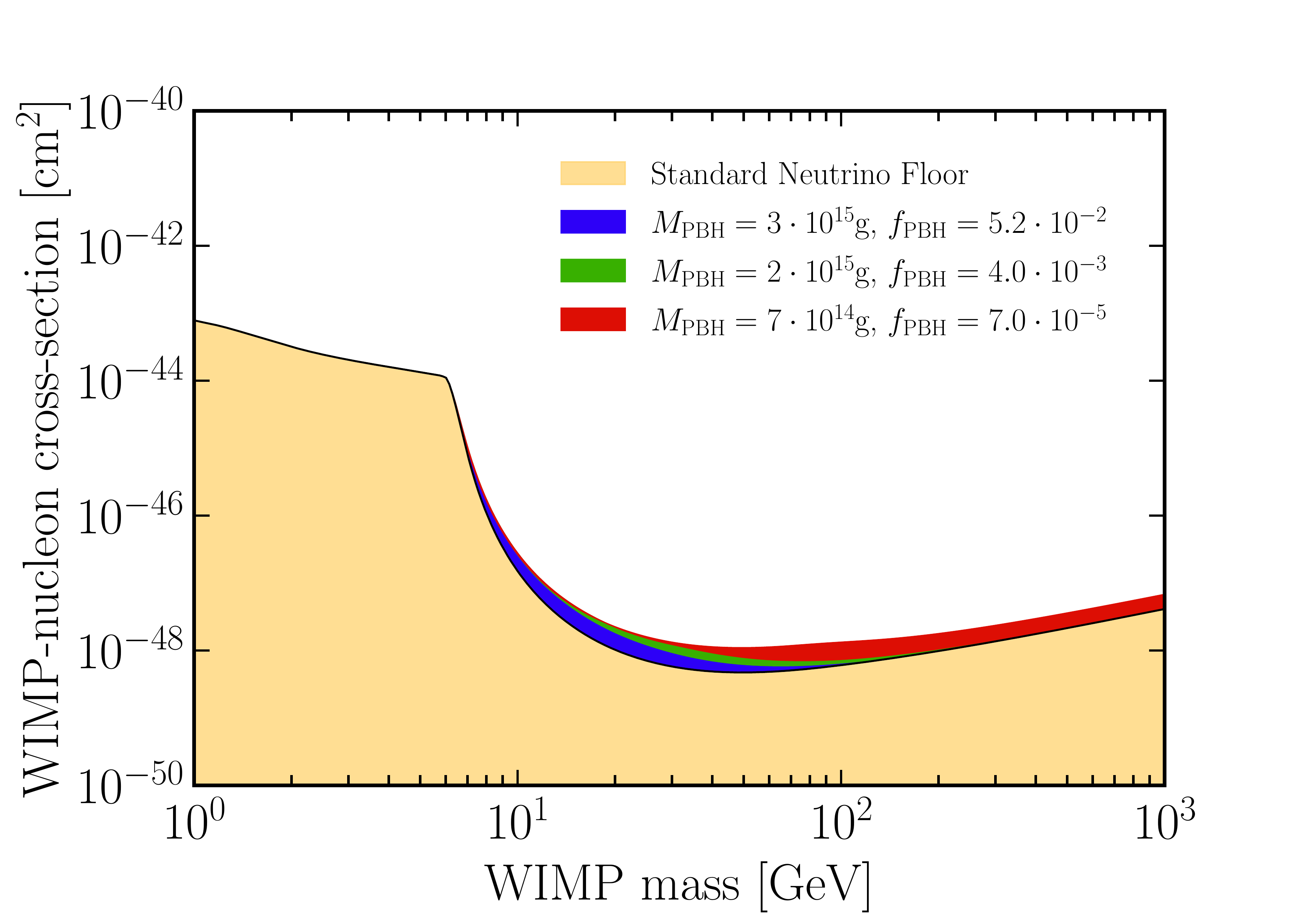}
\vspace*{-0.55cm}
    \caption{{\bf Impact of PBHs on the Neutrino floor.} The black contour delimiting
  the yellow region represents the ordinary neutrino floor, while the upper border
  of the colored bands correspond to the modifications induced by neutrinos from PBHs
   with masses and DM fractions in the legend.  These benchmark values lie on the
90\% C.L. exclusion curve obtainable from a liquid xenon experiment with 200 t yr exposure
(corresponding to the black stars in Fig.~\ref{fig:upper_bounds}).\label{fig:nu_floor}}
\end{figure}

As illustrated in the previous sections, the neutrinos emitted by PBHs would lie on top
of such a familiar background. Therefore, the existence of even a minute fraction
of PBHs in the DM content would modify the neutrino floor, making it higher. Here we
quantify this effect by calculating the floor following the prescription of~\cite{Billard:2013qya}. For definiteness
and consistency with the previous sections, we consider the case of a xenon target nucleus and
adopt the same NFW profile as in Eq.~(\ref{eq:NFW}). Following~\cite{Lewin:1995rx}, we employ a galactic 
Maxwell-Boltzmann velocity distribution with most probable speed $v_0$ = 220 km/s, 
truncated at the escape velocity $v_{esc}$ = 544 km/s and boosted into the laboratory frame with $v_{lab}$ = 232 km/s. 
For a fixed value of the WIMP mass we calculate the cross-section $\sigma_{\chi n}$
that can be excluded at the 90\% confidence level (corresponding to 2.3 DM events) selecting 
the exposure which leads to 1 CE$\nu$NS count. Then, varying over the energy threshold $E_R^{\rm thr}$, 
we keep the value which minimizes the cross-section. 
By repeating such a procedure for a dense grid of WIMP masses in the range $[1,1000]$\,GeV,
we construct the contours shown in Fig.~\ref{fig:nu_floor}. The black upper border of the yellow region
corresponds to the well-known ordinary floor. The upper edge of the colored regions
corresponds to the case of PBHs with parameters reported in the legend, where for the sake of clarity, 
we use the same values and color convention adopted in Figs.~\ref{fig:nu_fluxes} and \ref{fig:event_rate}.

Here, an important remark is in order. For sufficiently high PBH masses
$M_{\rm PBH} \gtrsim 10^{21}{\rm g} \sim 10^{-12}M_\odot$ ($M_\odot$ being the solar mass), 
the existence of WIMPs as a dominant or subdominant component
of DM, is incompatible even with a small fraction of DM made of
PBHs~\cite{Lacki:2010zf,Boucenna:2017ghj,Adamek:2019gns,Bertone:2019vsk,Cai:2020fnq,Carr:2020mqm,Kadota:2020ahr,Tashiro:2021xnj}.
This occurs because WIMPs are accreted by the PBH potential well, forming spiked ultra-compact mini-halos,
whose existence can be excluded by the non-observation of the expected products of WIMPs self-annihilation.
However, for the low PBHs masses considered in our work, falling in the so-called evaporation range 
($M_{\rm PBH}  \lesssim 10^{16}{\rm g} \sim 10^{-17}M_\odot$), this phenomenon is negligible.
In addition, one should note that DM may be asymmetric~\cite{Kaplan:2009ag,Petraki:2013wwa,Zurek:2013wia},
and this may occur also for WIMPs~\cite{Graesser:2011wi}, in which case self-annihilation
is reduced (albeit not necessarily absent~\cite{Cohen:2009fz,Cai:2009ia,Baldes:2017gzw,Baldes:2017gzu}).
Therefore, we deem it interesting to consider how the neutrino floor would get modified by the presence of 
a minute DM fraction of PBHs having mass in the range considered in this work. 

{\bf \em Conclusions.}  We have explored a new avenue in constraining PBHs 
as Dark Matter with the neutrinos emitted in the Hawking radiation.
Specifically, we have pointed out that neutrinos from PBHs can interact via 
the coherent elastic neutrino nucleus scattering (CE$\nu$NS) 
in multi-ton DM direct detection experiments.  We have shown that with the high exposures envisaged for the next-generation facilities,
 it will be possible to set bounds on the fraction of DM composed of PBHs, improving the existing neutrino 
limits obtained with Super-Kamiokande. In addition, we have quantified how much a signal originating from 
PBHs would heighten the so-called ``neutrino floor'', the ultimate barrier towards  detection of
WIMPs as the dominant DM component. For definiteness we have focused our study on liquid xenon detectors such as
DARWIN~\cite{Aalbers:2016jon}, LZ~\cite{Akerib:2018lyp},  and XENONnT~\cite{Aprile:2020vtw}, but other targets such as liquid argon employed in DarkSide-20k~\cite{Aalseth:2017fik} and ARGO~\cite{Aalseth:2017fik}, or archeological lead in
RES-NOVA~\cite{Pattavina:2020cqc}, should offer a similar opportunity provided that very high exposures are reached
(see~\cite{Billard:2021uyg} for an extensive review of the experimental program of the direct detection facilities).
Finally, we underline that, in the context of PBHs searches, the direct DM experiments would 
rather operate as indirect DM observatories.  From this perspective,  our study lends further support 
to the emerging role of such underground facilities as multi-purpose low-energy neutrino telescopes
complementary to their high-energy ``ordinary'' counterparts, IceCube and KM3NeT.\\

\section*{Acknowledgments}
We thank B. Dasgupta  and D. Montanino for helpful discussions.
We acknowledge partial support by the research grant number 
2017W4HA7S ``NAT-NET: Neutrino and Astroparticle Theory Network'' 
under the program PRIN 2017 funded by the 
Italian Ministero dell'Istruzione, dell'Universit\`a e della Ricerca (MIUR) 
and by the research project {\em TAsP} funded 
by the Instituto Nazionale di Fisica Nucleare (INFN).

\bibliographystyle{JHEP_new}
\bibliography{PBHs-References_2020}

\end{document}